\documentclass[conference]{IEEEtran}
\IEEEoverridecommandlockouts
\usepackage{cite}
\usepackage{amsmath,amssymb,amsfonts}
\usepackage{algorithmic}
\usepackage{graphicx}
\usepackage{textcomp}
\usepackage[dvipsnames]{xcolor} 
\usepackage[hyphens]{url}
\usepackage{marvosym}

\def\BibTeX{{\rm B\kern-.05em{\sc i\kern-.025em b}\kern-.08em
    T\kern-.1667em\lower.7ex\hbox{E}\kern-.125emX}}

\usepackage{color}
\usepackage{soul}
\usepackage{multirow}
\usepackage{makecell, rotating}
\usepackage{subfigure}
\usepackage{enumerate}
\usepackage{enumitem}
\usepackage{balance}
\usepackage[linesnumbered,ruled,lined]{algorithm2e}
\usepackage{threeparttable}

\usepackage{listings}
\usepackage{multicol}
\usepackage{tcolorbox}
\usepackage{tikz}
\usetikzlibrary{tikzmark}

\newcommand{\RebuttalChange}[1]{\textcolor{black}{#1}}

\newcommand{\SolutionName}{\emph{Taiyi}}
\newcommand{\Tofill}[1]{\textcolor{black}{#1}}

\title{Taiyi: A high-performance CKKS accelerator for Practical Fully Homomorphic Encryption}

\author{
    
        Shengyu Fan\IEEEauthorrefmark{1}\IEEEauthorrefmark{2}, 
        Xianglong Deng\IEEEauthorrefmark{1}\IEEEauthorrefmark{2}, 
        Zhuoyu Tian\IEEEauthorrefmark{1}\IEEEauthorrefmark{2}, 
        Zhicheng Hu\IEEEauthorrefmark{3},  \\
        Liang Chang\IEEEauthorrefmark{3}, 
        Rui Hou\IEEEauthorrefmark{1}\IEEEauthorrefmark{2}, 
        Dan Meng\IEEEauthorrefmark{1}\IEEEauthorrefmark{2}, 
        Mingzhe Zhang\IEEEauthorrefmark{1}\IEEEauthorrefmark{2}\textsuperscript{\Letter}\thanks{\textbf{Corresponding Author:} Mingzhe Zhang (zhangmingzhe@iie.ac.cn).
        }\\
        \IEEEauthorrefmark{1}Key Laboratory of Cyberspace Security Defense, Institute of Information Engineering, CAS, Beijing, China. \\
        \IEEEauthorrefmark{2} School of Cyber Security, University of Chinese Academy of Sciences, Beijing, China. \\
        \IEEEauthorrefmark{3} University of Electronic Science and Technology of China. \\
}

\begin{document}
\maketitle
\thispagestyle{plain}
\pagestyle{plain}


\begin{abstract}
Fully Homomorphic Encryption (FHE), a novel cryptographic theory enabling computation directly on ciphertext data, offers significant security benefits but is hampered by substantial performance overhead. In recent years, a series of accelerator designs have significantly enhanced the performance of FHE applications, bringing them closer to real-world applicability. However, these accelerators face challenges related to large on-chip memory and area. Additionally, FHE algorithms undergo rapid development, rendering the previous accelerator designs less perfectly adapted to the evolving landscape of optimized FHE applications.

In this paper, we conducted a detailed analysis of existing applications with the new FHE method, making two key observations: 1) the bottleneck of FHE applications shifts from NTT to the inner-product operation, and 2) the optimal \(\alpha\) of \texttt{KeySwitch} changes with the decrease in multiplicative level. Based on these observations, we designed an accelerator named \SolutionName, which includes specific hardware for the inner-product operation and optimizes the NTT and BConv operations through algorithmic derivation. A comparative evaluation of \SolutionName~against previous state-of-the-art designs reveals an average performance improvement of \RebuttalChange{1.5\(\times\)} and reduces the area overhead by 15.7\%.
\end{abstract}

\section{Introduction}\label{sec:intro}
Fully Homomorphic Encryption (FHE) enables direct computation on encrypted data, offering a powerful means to safeguard private information. Various FHE schemes have been proposed to cater to different application scenarios, including BGV~\cite{bgv}, BFV~\cite{BFV-1,BFV-2,BFV-3}, CKKS~\cite{ckks}, and TFHE~\cite{tfhe,lou2020glyph}. Among these, CKKS allows for computations on complex numbers in real-world applications and supports parallel processing of many slots~\cite{boemer2019ngraph,chen2019efficient,lou2021hemet}. Consequently, CKKS holds significant promise for the practical application of FHE in real-world scenarios.
For the CKKS scheme, the initial step involves encoding a vector containing multiple complex values into plaintext using the FHE compiler, and then the plaintext undergoes encryption.
Once encrypted, CKKS allows for direct multiplication and other linear arithmetic operations on the ciphertexts. 
The decryption and decoding phase is where the ciphertext is ultimately decrypted, and the outputs of this phase are the ultimately computed results.

The CKKS strategy, while powerful, faces significant computational overhead, posing challenges to the speed of application execution and hindering the practical deployment of FHE applications. Various methods have been proposed to address this issue and enhance FHE application performance across different platforms. On GPU platforms, strategies include algorithm optimization, bandwidth optimization, and the use of Tensor Cores to reduce the execution time of FHE applications~\cite{tensorfhe,100x}. Similarly, on FPGA platforms, methods involve optimizing dataflow and implementing modular fusion techniques to enhance the efficiency of FHE operations~\cite{poseidon,heax}. Despite these advancements, a substantial performance gap still exists, impeding the realization of FHE applications.
Consequently, domain-specific accelerators have been introduced to narrow this performance gap in real-world applications. These accelerators deploy specialized components for FHE arithmetic kernels, achieving high throughput and overall performance improvements for FHE applications~\cite{f1,ark,sharp,bts,craterlake}. Therefore, domain-specific accelerators contribute significantly to achieving optimal performance for FHE applications, a crucial step towards making FHE a practical reality in our lives in the future.


While existing FHE accelerators have showcased remarkable performance in executing FHE applications, the state-of-the-art implementation of \textit{SHARP} is particularly noteworthy. It has demonstrated significant performance improvements compared to CPU implementations and notable reductions in both area and on-chip storage when contrasted with previous accelerator solutions~\cite{sharp}. 
However, the FHE algorithms undergo rapid development, which presents new challenges for existing accelerators during execution. 

The recent algorithm-level optimizations for the CKKS scheme significantly enhance application performance by reducing the execution times of specific arithmetic kernels. Thus, the KLSS method decreases the execution times of the \texttt{NTT}, causing a significant change in the bottleneck~\cite{newkey,levelaware}.
Consequently, it needs to re-analyze both FHE applications and the parameter selection of the CKKS scheme, prompting a re-design of the FHE accelerator to align with the evolving landscape of FHE algorithm-level optimization.

This work focuses on optimizing performance for the KLSS-based method on the accelerator. We comprehensively analyze the KLSS method's characteristics, performing a runtime state analysis on the \textit{SHARP} accelerator. Key observations include a shift in crucial operations, with \textit{Inner Product} (IP) operations gaining prominence over the previously dominant Number Theoretic Transform (NTT). 
Despite the performance of NTT's continued importance, efficient execution hinges on addressing the challenges posed by IP operations, where a substantial accumulation of instructions in existing accelerators impedes performance.
Furthermore, our exploration underscores the suboptimal nature of conventional parameter selection with a fixed \(\alpha\) to the optimal \(\alpha\) dynamically changes with multiplication depth consumption.

To address the challenges inherent in KLSS-based methods, we present several optimizations to enhance the performance of FHE applications. Our contributions are as follows:
\begin{itemize}
    \item \RebuttalChange{\textbf{
    We bridge computer architecture design and timely cryptography} research. Prior FHE accelerator research focused on improving classical FHE scheme performance, assuming FHE algorithm changes are transparent to hardware. This paper demonstrates that specific FHE algorithm development poses challenges to existing accelerator architecture but also offers new optimization opportunities. Considering cryptography's timely progress, accelerator architecture optimization can significantly improve performance without increasing hardware overhead. This work highlights the importance of tracking cryptography's latest progress and may inspire new ideas for future FHE accelerators.
    }
    \item  \RebuttalChange{\textbf{We design dedicated hardware structures and acceleration schemes to address the new performance bottlenecks (\texttt{IP} and \texttt{BConv}) identified in the experimental analysis.} For \texttt{IP}s, we propose a hierarchical MAC structure (HPIP) and enhance data reuse through dataflow optimization. For \texttt{BConv}s, we propose the fusion method to reduce \texttt{BConv} in Recover-limbs stages and the ModDown.}
    \item \RebuttalChange{\textbf{We prove that the proper parameter selection significantly affects the CKKS performance and provides a new direction for the optimization of the FHE compiler.} We propose a dynamic parameter selection scheme and then experimentally analyze the impact of various parameter values. This will probably inspire more work on the optimization of compiler and accelerator for FHE workloads.}
    \item \textbf{Comprehensive Evaluation of \SolutionName:} We offer a comprehensive evaluation of \SolutionName~from diverse perspectives. The assessment results underscore the efficacy of our accelerator design for KLSS-based FHE applications, showcasing the \RebuttalChange{\(1.5\times\)} performance improvement across various application scenarios on average.

\end{itemize}

These contributions collectively address critical aspects of KLSS-based FHE methods, spanning both hardware and algorithmic domains, focusing on achieving superior performance for the FHE applications.

\section{Background}
In this section, we introduce the background related to this paper. Moreover, the symbols related to this paper are indicated in Table~\ref{tab:ckksnotaion}.

\subsection{CKKS: A Practical and Promising FHE Scheme}


Cryptographic schemes play a crucial role in safeguarding real-world data within the privacy and information protection domain. However, the complex operations involved in CKKS pose a significant challenge, making integrating this scheme with practical applications difficult. This section provides a brief overview of the CKKS scheme hierarchy to help readers understand the different layers and their relationships.





The hierarchy of the CKKS is as follows.
\begin{itemize}
    \item \textit{BOOT} layer. It facilitates the recovery of the multiplicative level of ciphertexts through the \texttt{Bootstrapping} operation, directly interfacing with the \textit{HEOp} layer. The \texttt{Bootstrapping} operation involves multiple \texttt{HMULT} and \texttt{HROTATE} operations in the \textit{HEOp} layer to achieve homomorphic decoding and encoding.
    \item \textit{HEOp} layer. It encompasses fundamental operations used in CKKS, including \texttt{HMULT}, \texttt{HROTATE}, \texttt{HADD}, \texttt{PMULT}, and \texttt{PADD}. It serves as a pivotal component of the CKKS scheme, accessing the \textit{Arithmetic} layer for all operations and requiring access to the \textit{KSW} layer for \texttt{HMULT} and \texttt{HROTATE} operations.
    \item \textit{KSW} layer. It is dedicated to serving the layers above it and incurs significant computational cost due to executing \texttt{NTT}, \texttt{BCONV}, and \texttt{IP} operations.
    \item \textit{Arithmetic} layer, It houses all arithmetic kernels, with each kernel serving the \texttt{KeySwitch} operation and supporting basic arithmetic operations (\texttt{ADD} and \texttt{MUL}) for operations in the \textit{HEOp} layer.
\end{itemize}

Applications secured with CKKS interact directly with the \textit{BOOT} and \textit{HEOp} layers, where original application operations transform into linear operations in FHE, including \texttt{HMULT} and \texttt{CMULT}. It also incorporates the \texttt{Bootstrapping} operation to refresh the multiplicative level.

Based on the above description, the \textit{KSW} layer plays a pivotal role in the CKKS hierarchy, acting as the bridge between FHE operations and arithmetic kernels. This layer significantly impacts the overall performance of FHE applications. The prominent performance bottleneck in CKKS arises from its considerable computational overhead. To address this challenge, we introduce a specialized accelerator designed to optimize performance, emphasizing enhancing the efficiency of the \texttt{KeySwitch} operation and, consequently, improving overall application efficiency.


\subsection{CKKS parameters}
\label{sec:back:b}

\begin{table}[t]
\vspace{-5pt}
    \centering
    \caption{The Maximum $\log{PQ}$ and Corresponding Target Levels to Achieve 128-bit Security for Different Degrees.}\label{tab:security_pq}
    
    \begin{tabular}{ccccc}\hline
    $N$&$h$&$\log{PQ}$&$\lambda$&$L_\text{target}$\\ \hline
    $2^{15}$& 512 &782  & 134.4& 22\\
    $2^{16}$& 512 &1656 & 133.0& 45\\
    $2^{17}$& 512 &3276 & 134.5& 91\\
    $2^{18}$& 512 &6804& 128.3& 189
    \\ \hline
    \vspace{-25pt}
    \end{tabular}
\end{table}
CKKS is a homomorphic encryption scheme that relies on lattice cryptography. Hence it is influenced by parameters such as the lattice dimension \(N\) and modulus \(PQ\). The increment of \(N\) enhances the security level \(\lambda\), while the increment in \(PQ\) results in a reduction of \(\lambda\). In the CKKS scheme, \(P\) represents the special moduli product, and \(Q\) stands for the ciphertext moduli product, both of which are correlated with \(L\) and \(K\), (\(L + K = \frac{\log(PQ)}{\text{word length}}\)).
As shown in Table~\ref{tab:security_pq}, to achieve 128-bit security for FHE, we employ an estimation tool~\cite{sparseLWE} to calculate the maximum allowable value of $PQ$ for the most commonly used values of $N$. Additionally, $PQ$ is influenced by parameters such as $L$ and $K$, with $K$ is determined by $\lceil \frac{L+1}{d_{\text{num}}} \rceil$. Consequently, we adopt a word length of 36 bits\footnote{According to Ref.~\cite{sharp}, a 36-bit word length is considered the minimum requirement to meet the precision needs of FHE applications. Nevertheless, increasing word length becomes necessary in scenarios where precision requirements are heightened. However, the maximum level analysis method proposed in this paper remains applicable in such cases.}, as recommended in \cite{sharp}. This choice allows us to derive the target level $L_\text{target}$ from $PQ$, which is also equivalent to $L+K$. We establish a relationship between $d_{\text{num}}$ and maximum level,
\begin{equation}
\label{ldnum}
L = \lfloor \frac{d_{\text{num}}*L_\text{target}-1}{d_{\text{num}}+1}\rfloor
\end{equation}
It enables us to identify the optimal value of $d_{\text{num}}$ for various degrees of $N$ while ensuring a security level of $\lambda>128$.

\subsection{KLSS: A Breakthrough \texttt{KeySwitch} Method}
\label{sec:back:c}

\begin{algorithm}[t]
\caption{ Hybrid \texttt{KeySwitch} method}\label{alg:hybridksw}
\begin{algorithmic}[1]
    \REQUIRE $\textbf{c}$ = \{$\hat{\textbf{c}}_{0},...,\hat{\textbf{c}}_{\beta}$\}, M=\(Nlog_{N}\), L=\(\alpha\beta\)
    \colorbox{SkyBlue}{\parbox{\dimexpr0.9\linewidth-3\fboxsep}{
    \FOR{$i$ in 0 to $\beta$}
    \STATE $\hat{\textbf{c}}_{i} = \{\hat{c}_{i}^{0},...,\hat{c}_{i}^{L}\} \leftarrow \text{NTT}(\text{BConv}(c_{i}^{0},...,c_{i}^{\alpha}))$
    \ENDFOR
    \STATE  \hfill $\triangleright$ \textit{Decomposition} (O(\((M+\alpha)L\beta\)))}}
    \FOR{$j$ in 0 to 2}
    \colorbox{yellow}{\parbox{\dimexpr0.9\linewidth-3\fboxsep}{
    \FOR{$n$ in 0 to $\beta$}
        \STATE $\widetilde{\textbf{c}}^{j}_{m} += \textbf{\(\hat{c}\)}_{n} * \textbf{evk}^{j}_{mn}$
    \ENDFOR
    \STATE \hfill $\triangleright$ (\textit{IP} O(\(\beta LN\)))}}
    \colorbox{Lavender}{\parbox{\dimexpr0.9\linewidth-3\fboxsep}{
        \STATE $\hat{\textbf{c}}^{j} = \{\widetilde{c}_{i}^{0},...,\widetilde{c}_{i}^{L}\} \leftarrow \text{INTT}(\widetilde{\textbf{c}}^{j}_{m})$
    \STATE \hfill $\triangleright$ \textit{Recover-form} (O(\(ML\)))}}
     \ENDFOR
    \RETURN \textbf{ModDown(\(\hat{\text{\textbf{c}}}^{0},\hat{\text{\textbf{c}}}^{1}\))}
\end{algorithmic}
\end{algorithm}

\RebuttalChange{
\texttt{KeySwitch} operation significantly impacts the runtime of FHE applications, constituting the majority of the overall execution time~\cite{newkey,f1,ark,han2020better}. Ref.~\cite{han2020better} introduced a hybrid \texttt{KeySwitch} method to balance $P$ and $Q$ for minimal execution times in arithmetic kernels, as shown in Alg.~\ref{alg:hybridksw}. However, this method remains time-consuming, particularly during \texttt{NTT} operations, leading to significant delays in overall FHE application execution. To enhance the \texttt{KeySwitch} operation's performance further, a new method named KLSS, illustrated in Alg.~\ref{alg:klss}, focuses on reducing \texttt{NTT} execution times through RNS-based gadget decomposition~\cite{newkey}. This approach computes the same formula differently from the previous KeySwitch method, avoiding additional noise growth during computation. Consequently, KLSS can be directly applied to various FHE applications, achieving further performance improvement without introducing additional noise impact.
}
As shown in Alg.~\ref{alg:klss}, KLSS consists of four main steps: \textit{Gadget-decomp}, \textit{Inner-Product~(IP)}, \textit{Recover-Limbs}, and \textit{ModDown}.

\RebuttalChange{
As shown in Alg.~\ref{alg:klss}, the first step (\textit{Gadget-decomp}) is performed using the \texttt{BConv} and \texttt{NTT} operations. All limbs are grouped into \(\beta\) groups, each containing \(\alpha\) limbs. Compared to the previous \texttt{KeySwitch} method shown in Alg.~\ref{alg:hybridksw}, this step significantly reduces the number of \texttt{NTT}-related operations from \(\alpha \beta^{2}\) to \(\beta \alpha^{'}\) times by grouping ciphertext into small groups. Additionally, it alters the scale for each \textit{BConv} operation from \(\alpha\beta\) to \(\alpha^{'}\). As a result, this step substantially reduces the computational overhead, leading to improved performance.
}

\RebuttalChange{
The second step involves \texttt{IP} operations, which multiplies a vector of polynomial groups by a matrix of polynomial groups. Compared to the hybrid Keyswitch method, this step significantly increases computational complexity from \(\alpha\beta^{2}\) to \(\beta\widetilde{\beta}(\alpha^{'})^{2}\), becoming the primary bottleneck of the entire \texttt{KeySwitch} operation. Efficiently executing the \texttt{IP} operation is crucial for improving \texttt{KeySwitch} performance.
}



\RebuttalChange{
The \textit{Recover-Limbs} stage is responsible for restoring the number of limbs for the ciphertext. Each group executes a \texttt{BConv} operation to convert the limbs back into the corresponding levels of the original limbs. This step closely resembles the hybrid \texttt{KeySwitch} operation and also involves \texttt{NTT} and \texttt{BConv} operations. However, in comparison to the previous method, the \textit{Recover-Limbs} stage is an additional step, which increases the number of \texttt{BConv} operations by \(\widetilde{\beta}\) times, each converting from \(\alpha^{'}\) limbs to \(\alpha\) limbs.
}

\begin{algorithm}[t]
\caption{KLSS \texttt{KeySwitch} method}\label{alg:klss}
\begin{algorithmic}[1]
    \REQUIRE $\textbf{c}$ = \{$\hat{\textbf{c}}_{0},...,\hat{\textbf{c}}_{\beta}$\}, M=\(Nlog_{N}\)
    \colorbox{SkyBlue}{\parbox{\dimexpr0.9\linewidth-3\fboxsep}{
    \FOR{$i$ in 0 to $\beta$}
    \STATE $\hat{\textbf{c}}_{i} = \{\hat{c}_{i}^{0},...,\hat{c}_{i}^{\alpha'}\} \leftarrow \text{NTT}(\text{BConv}(c_{i}^{0},...,c_{i}^{\alpha}))$
    \ENDFOR
    \STATE  \hfill $\triangleright$ \textit{Gadget-decomp} (O(\((M+\alpha)\alpha^{'}\beta\)))}}
    \FOR{$j$ in 0 to 2}
    \FOR{$m$ in 0 to $\widetilde{\beta}$}
    \colorbox{yellow}{\parbox{\dimexpr0.9\linewidth-3\fboxsep}{
    \FOR{$n$ in 0 to $\beta$}
        \STATE $\widetilde{\textbf{c}}^{j}_{m} += \textbf{\(\hat{c}\)}_{n} * \textbf{evk}^{j}_{mn}$
    \ENDFOR
    \STATE \hfill $\triangleright$ (\textit{IP} O(\(\beta\widetilde{\beta}\alpha^{'}N\)))}}
    \colorbox{Lavender}{\parbox{\dimexpr0.9\linewidth-3\fboxsep}{
        \STATE $\hat{\textbf{c}}^{j}_{i} = \{\widetilde{c}_{i}^{0},...,\widetilde{c}_{i}^{\alpha}\} \leftarrow \text{BConv}(\text{INTT}(\widetilde{\textbf{c}}^{j}_{m}))$
    \STATE \hfill $\triangleright$ \textit{Recover-Limbs} (O(\((M+\alpha)\alpha^{'}\widetilde{\beta}\)))}}
    \ENDFOR
     \ENDFOR
       \RETURN \textbf{ModDown(\(\hat{\text{\textbf{c}}}^{0},\hat{\text{\textbf{c}}}^{1}\))}
\end{algorithmic}
\end{algorithm}


In summary, the KLSS method significantly improves the performance of \texttt{KeySwitch} by reducing the number of \texttt{NTT} operations. However, it also increases the computational complexity of the \texttt{BConv} and \texttt{IP} operations. \RebuttalChange{In this paper, we conducted a detailed runtime analysis on the new algorithm and designed specific hardware and algorithm optimizations. These optimizations led to a reduction in the overhead of \texttt{BConv} and \texttt{IP}.}



\subsection{Four-Step NTT algorithm}

Previous designs prioritized NTT operation for overall FHE application execution~\cite{f1,ark,bts,sharp,poseidon}. The 4-step NTT method meets vector requirements and enhances batch processing. While enhancing NTTU, these methods added complexities, demanding a comprehensive understanding of the algorithmic process and specific \textit{twiddle-factor} requirements. This section offers an overview of 4-step NTT methods, focusing on algorithmic aspects.

\begin{equation}
\label{nttnormal}
X_{k} = \sum^{N-1}_{n=0}x_{n}W_{2N}^{2nk+n}
\end{equation}

Based on the standard NTT formula given in Eq.~\ref{nttnormal}, we can infer the formula for the four-step NTT. Thus, the formula is as follows:

\begin{equation}
\label{4nttbase1}
X_{k_{1}+N_{1}k_{2}} = \sum^{N_{2}-1}_{n_{2}=0}\sum^{N_{1}-1}_{n_{1}=0}x_{n_{1}N_{2}+n_{2}}W_{2N}^{(n_{1}N_{2}+n_{2})(2(k_{1}+N_{1}k_{2})+1)}
\end{equation}
where,
\begin{equation}
\label{4nttbas2}
\begin{aligned}
    n &= n_{1}N_{2}+n_{2} \\
    k &= k_{1}+k_{2}N_{2}
\end{aligned}
\end{equation}
Now, if we simplify the exponential, and we can get:
\begin{equation}
\label{4nttbas3}
\begin{aligned}
&W_{2N}^{(n_{1}N_{2}+n_{2})(2(k_{1}+N_{1}k_{2})+1)}\\
&= W_{2N}^{2k_{1}n_{2}+n_{2}+(2k_{1}+1)n_{1}N_{2}+2k_{2}n_{1}N+2N_{1}k_{2}n_{2}}\\
&=W_{2N_{1}}^{2k_{1}n_{1}+n_{1}}W_{2N_{2}}^{k_{2}n_{2}}\cdot 1\cdot W_{2N}^{2k_{1}n_{2}+n_{2}}
\end{aligned}
\end{equation}
Therefore, the four-step formula is summarized as follows. 
\begin{equation}
\label{4ntt}
\begin{aligned}
&X_{k_{1}+N_{1}k_{2}} \\ 
&=\sum^{N_{2}-1}_{n_{2}=0}\sum^{N_{1}-1}_{n_{1}=0}x_{n_{1}N_{2}+n_{2}}
W_{2N_{1}}^{2k_{1}n_{1}+n_{1}}W_{2N_{2}}^{k_{2}n_{2}}W_{2N}^{2k_{1}n_{2}+n_{2}}
\end{aligned}
\end{equation}

Eq.~\ref{4ntt} outlines the NTT operation's four steps: \(N_{1}\)-NTT, Transpose, Hadamard Product, and \(N_{2}\)-NTT. In the \(N_{1}\)-NTT operation, the twiddle-factor for the \(i\)-th Butterfly Unit in each stage (\(s\)) is \(W_{2N_{1}}^{2^{s}+i}\). For the \(N_{2}\)-NTT operation, the twiddle-factor for the \(i\)-th Butterfly Unit is \(W_{2N_{2}}^{i}\), and consistent across all NTT stages.

\RebuttalChange{Moreover, the previous accelerator optimized the 4-step NTT method into a 10-step process to reduce global communication during execution~\cite{sharp}. However, this introduces additional overhead due to changing requirements for the twiddle factor, leading to increased pipeline stalls. In this paper, we introduce a new NTT method based on the 4-step NTT through algorithmic inference, significantly mitigating the additional overhead of the 10-step NTT.}

\begin{table}[t]
    \centering
    \caption{Symbols and Notions used in this paper.}\label{tab:ckksnotaion}
    \vspace{-5pt}
    \begin{tabular}{lll}\hline
    \textbf{Symbol} & \textbf{Definition} \\ \hline
    \(\lambda\)& Security parameters for the FHE. \\
    Q& The product of all prime modulus values, denoted by $\prod_{i=0}^{L}q_i$.\\ 
    P& The product of special (prime) moduli, denoted by $\prod_{k=0}^{K}p_k$. \\
   \RebuttalChange{T}& \RebuttalChange{The product of prime moduli, mutually exclusive with P.} \\
    PQ & The maximum moduli of the lattice. \\
    L& The maximum multiplicative level, which is related to Q. \\ 
    N& Degree of a polynomial and the lattice dimension.\\ 
    $l$& Current multiplicative level \\ 
    \multirow{1}{*}{$d_\text{num}$}& The number of digits in the switching key.~\cite{han2020better}\\
   \multirow{1}{*}{$\alpha$} & $\alpha$=$\lceil\frac{L+1}{d_\text{num}}\rceil$,~Number of limbs that comprise  \\ 
   &a single digit in the \texttt{KeySwitch} decomposition~(K=\(\alpha\)). \\
   \multirow{1}{*}{$\beta$} &  $\beta$=$\lceil\frac{l+1}{\alpha}\rceil$. $l$ limbs polynomials are split into \(\beta\) groups.\\ &this number of digits during base decomposition.\\ 
   \(\alpha^{'}\)& The number of limbs in \(R^{T}\), which used in KLSS-method. \\
   \(\widetilde{\beta}\) & The groups of limbs used in \texttt{IP}, which is \(\lceil\frac{l+\alpha+1}{\alpha^{'}}\rceil\). \\
   
    \texttt{fftIter}& Multiplicative depth of a linear transform in bootstrapping.\\ \hline
    \end{tabular}
     \vspace{-10pt}
\end{table}

\subsection{Overview of Previous FHE Accelerators}


To enhance the performance of CKKS, previous efforts involved designing domain-specific accelerators with customized functional units optimized for specific arithmetic kernels~\cite{f1,bts,craterlake,ark,sharp}. We briefly introduce the functional previous design for the main component as follows:
\begin{itemize}
    \item \texttt{NTT:} \RebuttalChange{\texttt{NTT} component executes \textit{NTT}-related kernels, including \textit{Butterfly} and \textit{transpose} units. 
    The \textit{ten-step} method, considered state-of-the-art, reduces communication within the \texttt{NTT} component by optimizing the algorithm~\cite{sharp}. However, pipeline stalls still occurred due to necessary twiddle-factor modifications during execution. Our algorithm optimization aims to eliminate these stalls to further improve performance.}
    \item \texttt{BConv:} The \textit{Basis Conversion} operation can be divided into two steps: the first step involves element-wise multiplication, while the second step consists of a matrix-vector multiplication operation. For the second step, the \texttt{BConv} component is designed to facilitate parallel computing with multiple MAC units. Consequently, this stage can be rapidly executed in parallel by leveraging a large number of MAC computing units~\cite{ark,sharp}.
    \item \texttt{AUTOU:} \RebuttalChange{The \textit{Automorphism} operation involves moving data based on the mapping index, which can be an irregular permutation among all $N$ coefficients. This operation is particularly useful for encoded ciphertexts, as it allows for coefficient rotation without decryption. The rotation is performed by the \texttt{AUTOU} unit, which consists of several MUX units arranged in eight stages. This component includes internal permutation logic to ensure high performance during related operations~\cite{ark}.}
    \item \texttt{EWE:} \RebuttalChange{Element-Wise Engine (\texttt{EWE}) component processes various element-wise operations, including the Hadamard Product used by Basis Conversion, ADD, or MUL operations. It also serves the IP operation in the previous accelerator, as the execution time of the IP operation is relatively low in the traditional KeySwitch method.}
\end{itemize}

\RebuttalChange{While previous efforts have notably improved CKKS application performance, the optimization of the KLSS method has brought significant changes to FHE application execution. While previous accelerators performed well, they faced challenges when executing the KLSS-based \texttt{KeySwitch}. In this paper, we conduct a detailed analysis of KLSS-based CKKS applications and optimize the components accordingly to align with the rapid development of the CKKS algorithm.}


\section{Motivation}
In this section, we conduct an in-depth analysis of its computational distinctions compared to the previous method~\cite{han2020better}, we achieve some observations and opportunities for designing an accelerator with the KLSS-based method.


\subsection{Contrasting Computation Workload and Memory Requirements}
\label{sec:motiv:a}
The KLSS method is a novel \texttt{KeySwitch} method, significantly improving the FHE application performance by reducing the execution times of \texttt{NTT}~\cite{newkey}.
To discern the difference between KLSS and the previous methods from a hardware perspective, we comprehensively analyze computational complexity breakdowns and memory requirements for both \texttt{KeySwitch} methods. 

\begin{figure}[t]
    \centering
    \includegraphics[width=1\linewidth]{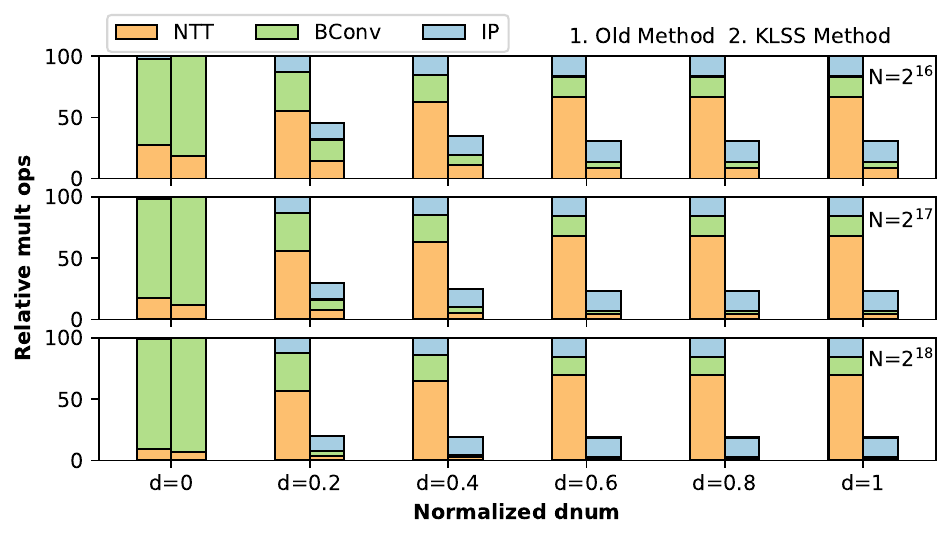}
    \vspace{-15pt}
    \caption{Computational arithmetic multiplication number breakdown. Results are measured for all possible values of \(d_\text{num}\) and N. 
    }
    \vspace{-10pt}
    \label{fig:mulops}
\end{figure}

\begin{table}[t]
    \centering
        \caption{\RebuttalChange{For the small \texttt{dnum} parameter, the KLSS method reduces the total number of multiplication operations compared to the previous KeySwitch method.}}
        \vspace{-5pt}
    \label{tab:reduction}
    \begin{tabular}{cccc}
    \hline
	&N=16	&N=17	&N=18 \\ \hline
dnum=2	&1.06&	1.12&	1.17\\ 
dnum=3	&9.9\%&	5.7\%&	6.4\%\\ 
dnum=4	&23.8\%&	20.6\%&	17.3\%\\ 
dnum=5	&34.1&	32.3\%&	29.9\%\\  \hline
    \end{tabular}
    \vspace{-15pt}
\end{table}

\begin{figure}[t]
    \centering
    \includegraphics[width=\linewidth]{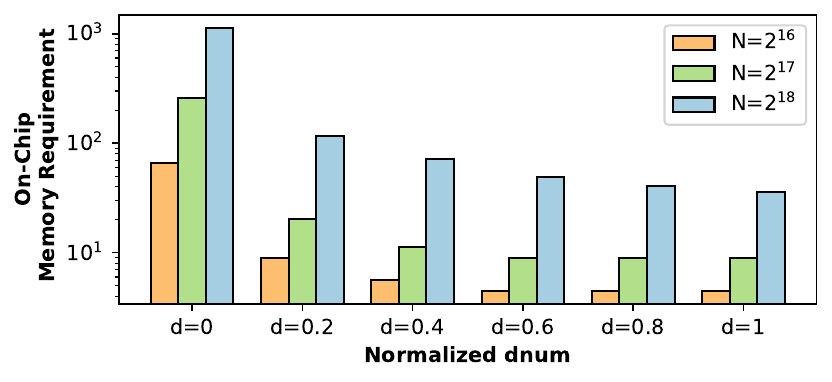}
    \vspace{-25pt}
    \caption{The On-Chip Memory Requirement for the KLSS-based \texttt{KeySwitch} method for \texttt{IP} operation.}
    \vspace{-10pt}
    \label{fig:memreq}
\end{figure}

As shown in Figure~\ref{fig:mulops} and Tabel~\ref{tab:reduction}\footnote{Normalized \(d\) = 0 means \(d_\text{num}\) = 1, and normalized \(d\) = 1 means \(d_\text{num}\) = maximum for each \(N\). Figure~\ref{fig:memreq} and Table~\ref{tab:param} also use this description.}, a stark contrast emerges between the KLSS method and its predecessor in the proportional relationships of \texttt{NTT}, \texttt{BConv}, and \texttt{IP} operations. 
KLSS achieves a remarkable reduction in the number of multiplication times for \texttt{NTT} by 59.6\%, leading to a \Tofill{$3.6\times$} performance enhancement.
However, \texttt{IP} and \texttt{BConv} operations within KLSS show are also increased, where \texttt{IP} surging by \Tofill{44.3\%} and \texttt{BConv} experiencing a noteworthy average proportion increase. 
The number of multiplication operations in \texttt{IP} increases on average by \Tofill{3.2\%}, and \texttt{BConv} experiences a rise of \Tofill{19.3\%} in some cases. 
Compared to the previous method, KLSS presents a notably distinct distribution of computational complexity. Traditionally, emphasizing the optimization of \texttt{NTT} performance in prior hardware accelerators(\cite{bts,ark,craterlake,f1,sharp}) may not yield substantial benefits in FHE acceleration when employing the KLSS method.

As shown in Figure~\ref{fig:memreq}, we note the difference in memory requirements for the \texttt{IP} operation in KLSS compared to the previous method.
Each group needs to concurrently compute $\beta\times \alpha^{'}$ ciphertexts in the KLSS-based method, which requires an average of \Tofill{102.9}~MB of memory. 
In contrast, the previous accelerator processes only $\alpha$ ciphertexts per group.
This results in potential challenges related to re-fetching the same ciphertext during the \texttt{IP} operation, leading to additional computational and memory overhead. 

The observations indicate a significant challenge for the previous accelerator when executing FHE applications optimized with the KLSS method. Initially designed with a focus on powerful \texttt{NTT} units, the prior accelerator faces a mismatch in computational complexity, since \texttt{IP} becomes the most complex operation. The \texttt{IP} operation incurs additional computational and memory overhead on the previous accelerator due to its sequential computation at the ciphertext group level, lacking parallel processing capabilities. Consequently, there is a need to reconsider FHE accelerator design, particularly in light of the emerging \texttt{KeySwitch} method, to achieve further performance enhancements for FHE applications.



\subsection{Performance Analysis and Optimal CKKS Parameter Selection}
\label{sec:motiv:b}

\begin{table}[t]
 \caption{$T_\text{mult,a/slot}$ are measured for all possible integer $d_\text{num}$ values for each $N$ and satisfy the security requirement of \(\lambda >\)128.}\vspace{-5pt}
    \centering
    \begin{tabular}{cccc}
    \hline
&N=\(2^{16}\)&N=\(2^{17}\)&N=1\(2^{18}\) \\ \hline
d=0    &28.97&190.22&1665.57\\ 
d=0.2&18.76&97.35&639.20\\ 
d=0.4&20.82&119.80&897.82\\
d=0.6&25.43&157.40&1246.74\\ 
d=0.8&25.43&157.40&1246.74\\ 
d=1   &25.43&157.40&1246.74\\ \hline
    \end{tabular}
    \label{tab:param}
    \vspace{-15pt}
\end{table}

The FHE application's performance is intricately related to its parameters. Previous research introduced the \textit{amortized multiplication time per slot} ($T_\text{mult,a/slot}$) as a metric to identify optimal parameters~\cite{100x}, which is defined as:
\[ T_{\text{mult,a/slot}} = \frac{T_\text{boot} + \sum_{l=1}^{L-L_\text{boot}}T_\text{mult}^{(l)}}{L-L_{\text{boot}}} \cdot \frac{2}{N} \]
where $T_\text{boot}$ is bootstrapping time and $T_\text{mult,a/slot}$ represents the time for \texttt{HMult} at level $l$. Both $T_\text{boot}$ and $T_\text{mult}$ impact the \texttt{KeySwitch} operation. The unique computational behavior of the KLSS method necessitates a re-evaluation of CKKS parameters for optimal performance.

For \texttt{Bootstrapping}, our focus is on determining the \texttt{fftIter} parameter, representing the iterations for homomorphic DFT~\cite{chen2019improved}. To ensure optimal performance, we set \texttt{fftIter} to 6, aligning with recommendations from advanced research~\cite{mad,de2021does}. we calculate $L_\text{boot}$ as $2 \times \texttt{fftIter} + 9 = 21$, signifying that the maximum level must exceed 21 to meet security and multiplicative depth requirements.

As discussed in the section \ref{sec:back:b}, the security of FHE is intricately related to the parameters $N$ and $PQ$. As $N$ increases, the security level $\lambda$ is enhanced, whereas an increase in $PQ$ leads to a reduction in $\lambda$. A larger value of $PQ$ facilitates more multiplicative operations, thereby reducing the necessity for executing the \texttt{Bootstrapping} operation in FHE applications. These observations underscore the crucial role that parameter selection plays in achieving the desired security and performance characteristics.

\begin{figure}[t]
    \centering
    \includegraphics[width=\linewidth]{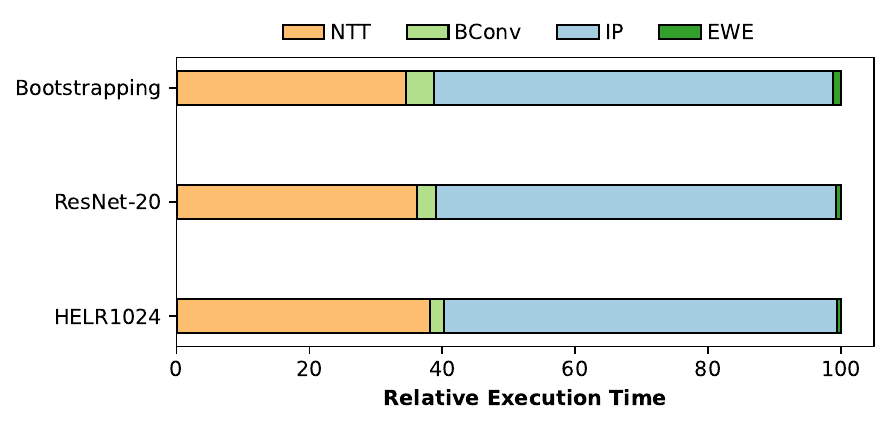}
    \vspace{-25pt}
    \caption{\texttt{KeySwitch} Execution Time Breakdown in FHE Application Using the KLSS-based method. \RebuttalChange{The \texttt{KeySwitch} execution time constitutes 73.7\%, 85.5\%, and 84.9\% of the total execution time for the respective cases.}}
    \vspace{-15pt}
    \label{fig:appbreak}
\end{figure}

As shown in Eq.~\ref{ldnum}, considering the relationship between \(L\) and \(d_{\text{num}}\), we evaluate the performance of \(T_{\text{mult,a/slot}}\) for different degrees\footnote{
To address pipeline overlap and component parallelization influences, we sequentially measure execution time for each compartment. These effects are contingent on computational component design, introducing a potential disparity between execution time and computational complexity, thus influencing parameter selection~\cite{bts,han2020better}.
}. Table~\ref{tab:param} presents \(T_{\text{mult,a/slot}}\) values for the FHE accelerator with the KLSS method. Thus, we can get the best performance for each \(N\), which achieves \Tofill{18.76}ns, \Tofill{97.35}ns, and \Tofill{639.2}ns for (\Tofill{\(2^{16}\), \(38\), \(6\)}), (\Tofill{\(2^{17}\), \(86\), \(18\)}) and (\Tofill{\(2^{18}\), \(194\), \(39\)}), respectively. Notably, \(N\) at \(2^{15}\) yields only one multiplicative level apart from \(L_{\text{boot}}\), potentially unsuitable for FHE with \texttt{Bootstrapping}. 
Besides, settings with \(N > 2^{16}\) demand excessive computational resources. 
Thus, our parameter used in this work is \(N = 2^{16}\).

In this paper, to design a more suitable FHE accelerator for the KLSS method, we illustrate our method using the CKKS instance with \(N\) = \(2^{16}\), \(L\) = \Tofill{38}, and \(d_{\text{num}}\) = \Tofill{6}.




\subsection{Real-world application breakdown analysis}
\label{sec:motiv:c}

To gain a more detailed understanding of the bottlenecks associated with the KLSS method in previous accelerators, we develop a simulator based on the SimFHE~\cite{mad}, and modeling the \textbf{SHARP} architecture for executing optimized FHE applications with the KLSS method.
As illustrated in Figure~\ref{fig:appbreak}, we observe that the \texttt{IP} operation constitutes the majority of the processing time in the overall application. The heightened computational workload of the \texttt{IP} operation can account for most of the execution time of the \texttt{EWE} component. Consequently, we can infer that the bottleneck of the FHE application has shifted from the \texttt{NTT} operation to the \texttt{IP} operation. The proportion of \texttt{NTT} operation only achieves \Tofill{36.3\%} on average, while the proportion of \texttt{IP} operation achieves \Tofill{59.8\%} on average and reaches 60.2\% for the ResNet-20 application. Our analysis suggests that the majority of \texttt{IP} instructions are constrained by the limited computation component of the \textbf{SHARP} accelerator. This limitation significantly impacts the execution efficiency, particularly as \texttt{IP} operations become a predominant factor in the overall workload. 


In summary, the previous FHE accelerator is not well-suited to the new changes in the FHE algorithm. 
Therefore, it is necessary to meet the high computational requirement of IP by re-designing the FHE accelerator.


\subsection{Diverging the best $\alpha$ for different multiplicative level}
\label{sec:motiv:d}
\begin{figure}[t]
    \centering
    \includegraphics[width=0.95\linewidth]{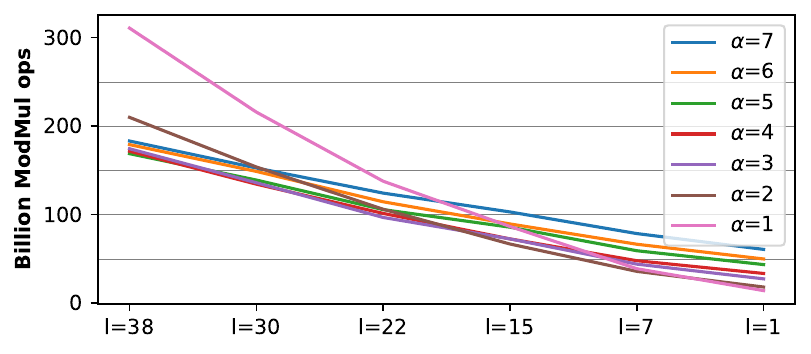}
    \vspace{-15pt}
    \caption{
    \RebuttalChange{The relationship between the number of ModMul operations and the decrease of multiplicative level~(\(l\)). \(\alpha\) is the number of single digits.}}
    \vspace{-15pt}
    \label{fig:level}
\end{figure}
As detailed in Section~\ref{sec:motiv:b}, the parameter \(d_{\text{num}}\) is fixed before FHE operations, determining the corresponding \(\alpha\) for the \texttt{KeySwitch} operation. Our analysis establishes the optimal \(d_{\text{num}}\) for security and maximal multiplicative level \(L\). However, \texttt{HMULT} and \texttt{CMULT} reduce the multiplicative level budget, affecting \texttt{NTT} and \texttt{Hadamard Product} execution times. Figure~\ref{fig:level} illustrates that, for KLSS, the optimal \(\alpha\) changes as \(l\) decreases. In real-world FHE workloads, where \(l\) decreases, a fixed \(\alpha\) may not yield optimal performance. 
To enhance performance for dynamic multiplicative levels, we propose dynamic \(\alpha\) adjustment. Prior studies~\cite{levelaware} also affirm that dynamic \(\alpha\) selection optimizes \texttt{KeySwitch} performance.
Thus, in this paper, we propose a static compiler to dynamic select \(\alpha\) for further improving the performance of KLSS-based FHE applications.

\subsection{Opportunities}
Based on the above observations, we conclude our opportunities to design the new FHE accelerator as follows:


\begin{itemize}
\item The KLSS-optimized FHE operation, as discussed in Sections~\ref{sec:motiv:a} and~\ref{sec:motiv:c}, significantly reduces the computational workload while increasing demands on the \texttt{IP} operation. This underscores the imperative need for dedicated hardware to support \texttt{IP}, along with the development of an optimized data flow to reduce memory requirements. Additionally, there is a necessity for further optimizing the \texttt{NTT} component due to its high overhead.

\item The \texttt{BConv} operation, as outlined in Sections~\ref{sec:motiv:c} and~\ref{sec:motiv:a}, experiences increased execution time with KLSS-based \texttt{KeySwitch}.
Algorithmic optimizations are essential to alleviate redundant \texttt{BConv} operations.


\item Sections~\ref{sec:motiv:b} and~\ref{sec:motiv:d} highlight the ability to select optimal values for \(L\) and \(d_\text{num}\) to meet security requirements. Developing a static compiler to further optimize FHE application performance is crucial to determining the ideal \(\alpha\) for each multiplicative level. This compiler should analyze consumption at the FHE application level and set the optimal \(\alpha\).
\end{itemize}

Based on the above opportunities, we propose a specific accelerator, named \SolutionName~, which improves the performance of KLSS-based FHE applications by algorithm optimization and hardware design. The following section briefly introduces our domain-specific architectures and algorithm optimizations.

\section{Design}

To improve the performance of KLSS-based FHE applications, we analyze hardware behavior and develop \SolutionName, an optimized accelerator. Focusing on critical aspects of the KLSS-based method, we design specific components, refine data flow, and introduce a compiler for customized instruction generation at various levels of multiplication.By implementing this comprehensive approach, the efficiency and overall performance of KLSS-based FHE applications are improved.

\subsection{Multi-step (I)NTT Architeture}
\label{sec:design:a}
While the prominence of NTT has diminished in KLSS-based FHE applications, it remains crucial, serving as a bridge between \texttt{BCONV} and \texttt{IP} operations. In this section, we introduce optimizations for the NTT operation through algorithm derivation.


In this section, we propose the Multi-step method to perform \texttt{NTT} operation, which is an extension of the four-step NTT method. Based on Eq.~\ref{4ntt}, the multi-step NTT comprises \(N_1\)-NTT and \(N_2\)-NTT, each with four steps, minimizing on-chip communication during computation. \RebuttalChange{Additionally, the new twiddle factor generation scheme eliminates the pipeline stall caused by updating the second-stage twiddle factor in Sharp~\cite{sharp}.} In \(N_1\)-NTT, the exponent of the \textit{Twiddle-factor} (\(W_{2N_{1}}^{2k_{1}n_{1}+n_{1}}\)) mirrors that of the standard NTT (\(W_{2N}^{2nk+n}\)), with the base changing from \(N\) to \(N_{1}\). However, the \textit{Twiddle-factor} for \(N_{2}\)-NTT differs from the previous operation. Subsequently, adhering to the basic formula, we will restate the formulation for the \(N_{2}\)-NTT operation.

\begin{equation}
\label{ntt2}
X_{k_{2}} = \sum^{N_{2}-1}_{n_{2}=0}x_{n_{2}}W_{2N_{2}}^{2n_{2}k_{2}}
\end{equation}
Therefore, the four-step NTT formula for \(N_{2}\) is as follows:
\begin{equation}
\label{ntt2four}
\begin{aligned}
&X_{k_{21}+N_{21}k_{22}} \\ 
&=\sum^{N_{22}-1}_{n_{22}=0}\sum^{N_{21}-1}_{n_{21}=0}x_{n_{21}N_{22}+n_{22}}
W_{2N_{2}}^{2(n_{21}N_{22}+n_{22})(k_{21}+N_{21}k_{22})} \\
&= \sum^{N_{22}-1}_{n_{22}=0}\sum^{N_{21}-1}_{n_{21}=0}x_{n_{21}N_{22}+n_{22}}W_{2N_{21}^{2n_{21}k_{21}}}W_{2N_{22}^{2n_{22}k_{22}}}W_{2N2}^{n_{22}k_{22}}
\end{aligned}
\end{equation}

In both the \(N_{21}\)-NTT and \(N_{22}\)-NTT operations, the \(i\)-th Butterfly Unit's twiddle-factor is constant, given by \(W_{2N_{2}}^{i}\). This twiddle factor remains unchanged across all stages of the NTT process. Consequently, the ten-step NTT method employs a fixed set of twiddle factors for every Butterfly Unit in each limb, facilitating the design of a hierarchical NTT architecture.

\begin{figure}[t]
    \centering
    \includegraphics[width=\linewidth]{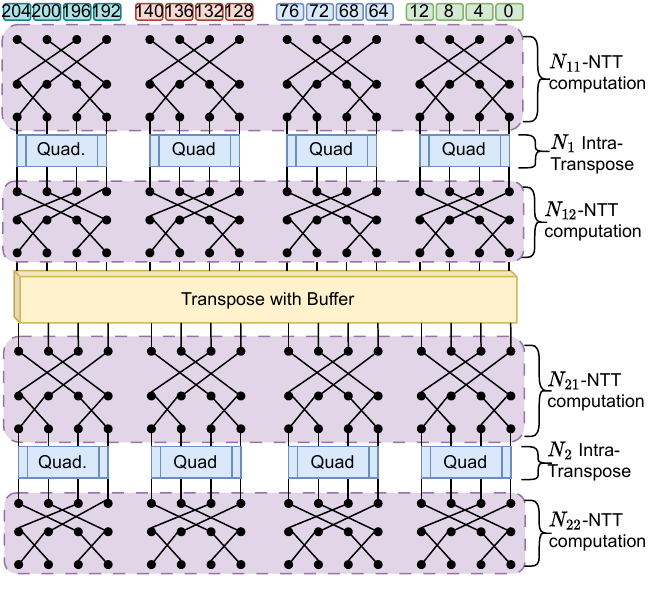}
    \vspace{-25pt}
    \caption{Organization of the NTTU and the data-map for input polynomial. Each NTTU serves 256 elements in one cycle, and \SolutionName~contains 16 lanes and each lane processes 16 elements on one cluster. For simplicity, here we take \textit{M}=4 for each NTTU group as an example.}
    \vspace{-15pt}
    \label{fig:nttworkflow}
\end{figure}

We can design the specific NTTU to accelerate the NTT-related computation based on the above algorithm understanding. As shown in Figure~\ref{fig:nttworkflow}, the workflow of NTTU can be organized into 7 stages, which are as follows.
\begin{itemize}
    \item \(N_{11}\)-NTT computation. During the computation of \(N_{11}\)-NTT, NTTU fetches multiple rows of one limb. Each lane processes one row by dividing it into \(N_{11}\) rows, where each row contains \(N_{22}\) elements. This ensures that the total number of elements in each row is equal to \(N_{1}\), where \(N_{11}\times N_{12}=N_{1}\). To handle 16 elements simultaneously, \SolutionName~utilizes 16 lanes. As a result, the hardware of the \(N_{11}\)-NTT undergoes a full-pipeline process for rows of limbs. Each lane has a butterfly layer depth of 4 and 8 butterfly units per layer.
    \item \(N_{1}\) Intra-Transpose. \SolutionName~conducts a transpose operation on the sub-rows of the limb for program execution legality. This is achieved using quadrant swap units (Quad.) and a Hadamard Product. The Quad units in each lane perform the transpose operation, featuring 18 quad units ranging from \(16\times16\) to \(2\times2\). Simultaneously, the Hadamard Product manages the multiplicative operation for each element, with the necessary twist factor generated by the OF-Twist units~\cite{ark}. For this stage, we implement the full-pipelined quad unit proposed in \cite{f1}.
    \item \(N_{12}\)-NTT computation. 
    This stage uses the same hardware configuration as \(N_{11}\)-NTT computation, but with a different twiddle factor, 
    utilizing \(W_{2N_{2}}^{i}\) as the fixed twiddle factor based on the index of the butterfly unit.
    \item Transpose with Buffer. In this stage, the limbs are transposed, organizing the values in row-wise. Subsequently, the above process performs \texttt{Hadamard Product} operations. Once all rows are stored in the buffer, the subsequent process reads them column-wise.
    \item \(N_{21}\)-NTT computation. Based on our previous algorithmic inference, we can deduce that this operation exhibits similar hardware behavior to both the \(N_{11}\)-NTT and \(N_{12}\)-NTT operations. The twiddle factor also remains consistent with the \(N_{12}\)-NTT operation, and the twiddle factor for each limb's computation is fixed.
    \item \(N_{2}\) Intra-Transpose. For this stage, the hardware behavior mirrors of \(N_{1}\) Intra-Transpose. The twist factor differs from the preceding operation, as it is based on \(W_{N_{2}}\).
    \item \(N_{22}\)-NTT computation. In this stage, the hardware structure, behavior, and twiddle factor remain consistent with \(N_{21}\)-NTT computation. All twiddle factors are fixed and do not vary based on our previous algorithm inference.
\end{itemize}

\begin{figure}[t]
    \centering
    \includegraphics[width=1\linewidth]{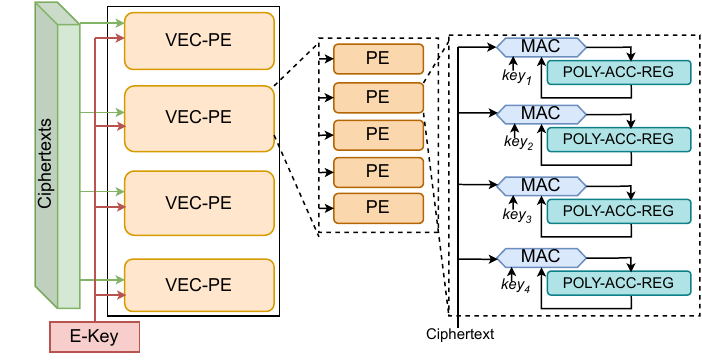}
    \vspace{-20pt}
    \caption{Illustration of the \textit{VEC-PE} array structue.}
    \label{fig:vecpearray}
    \vspace{-20pt}
\end{figure}

Our NTTU design minimizes on-chip communication and removes the requirement for additional twiddle factors generation by algorithmic inference and profound understanding. This design choice positions our NTTU to markedly improve the performance of the NTT operation.

\subsection{High Parallelism computation for \texttt{IP} operation.}

\label{sec:hpip}

\begin{figure}[t]
    \centering
    \includegraphics[width=0.9\linewidth]{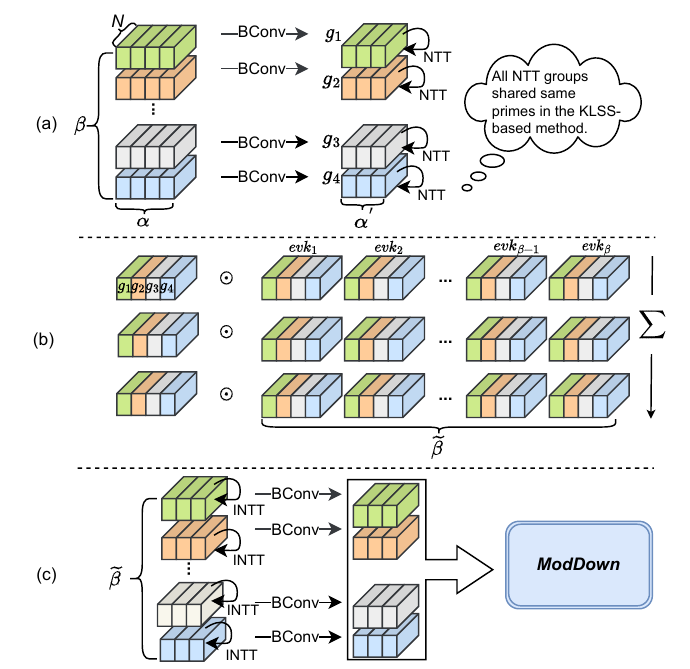}
    \caption{The workflow and data mapping of the KLSS-based \texttt{KeySwitch} method.
    (a) The operations organization of \textit{Gadget-decomp}.
    (b) \textit{IP} operations for the KLSS-based \texttt{KeySwitch}, and the data organization used in \SolutionName.
    (c) \textit{Recover-Limbs \text{and} ModDown} operations used in the KLSS-based \texttt{KeySwitch}.}
    \vspace{-15pt}
    \label{fig:kswnew}
\end{figure}

\RebuttalChange{As depicted in the Figure~\ref{fig:kswnew}, the data undergoes operation \textit{Gadget-decomp}~(Figure~\ref{fig:kswnew}(a)) before proceeding to operation \textit{IP}~(Figure~\ref{fig:kswnew}(b)) operation, followed by operations \textit{Recover-limbs} and \textit{ModDown}~(Figure~\ref{fig:kswnew}(c)). It was previously noted that operation \textit{IP} is the bottleneck of KLSS.}
Therefore, \SolutionName~introduces \textit{HP-IP}, a dedicated component designed for parallel \texttt{IP} operations. Comprising the \textit{VEC-PE} array and the \textit{E-Key Buffer}, \textit{HP-IP} executes element-wise multiplication using multiple MAC units, while the \textit{E-Key Buffer} stores and prefetches keys for \texttt{IP} operations. Subsequent sections provide detailed insights into each constituent of \textit{HP-IP}.

\subsubsection{\textit{VEC-PE} array}
As illustrated in Figure~\ref{fig:vecpearray}, the \textit{VEC-PE} array comprises \textit{M} VEC-PEs, each of which incorporates \textit{H} processing elements (PEs). In each cycle, every one of the 256 PE units located at the same horizontal position receives a data element~(36-bit used in this paper). These 256 data elements pertain to the same batch of polynomials and will undergo multiplication with the corresponding elements of the key. Subsequently, it is notable that the same PE will share these data elements.

\RebuttalChange{To efficiently calculate IP operations, we utilize data flow planning to reuse the input ciphertext during the IP calculation process.}
For each PE, consisting of 6 MAC units, following an output-stationary architecture, each MAC unit serves a distinct \textit{E-Key}, while all sharing the same prime and input limbs. The input limbs are directed into the PE. Within each MAC unit, the same batch of polynomials undergoes multiplication with the key's elements. Subsequently, the results are aggregated in the 128-bit POLY-ACC-REG.
To facilitate more efficient computations and reduce the need for frequent parameter switching during module operation, our design ensures that each MAC unit in the PE receives limbs from different groups at the same location. This approach significantly enhances the efficiency of \texttt{KeySwitch}.

\RebuttalChange{As depicted in Figure~\ref{fig:kswnew}(b), the dataflow of the VEC-PE involves performing vector-wise operations for PEs at the granularity of the ciphertext group. Each VEC-PE contains \textit{H} PEs.}
These PEs execute limbs from different groups in parallel; \(\text{PE}_{0}\) computes \(g_{0}\) to \(g_{5}\), while \(\text{PE}_{1}\) handles \(g_{6}\) to \(g_{11}\). It's worth noting that while increasing \textit{H} enhances performance, it also demands higher bandwidth. Therefore, we set \textit{H} to 4 to balance performance and bandwidth for \textit{HP-IP}. In cases where \(\beta\) is smaller than 4, our design allocates additional limbs from \(\widetilde{\beta}\) to different PEs to further enhance performance.

For \texttt{IP} operations in \SolutionName's \textit{HP-IP} component, we integrate 256 VEC-PE units to process all elements in one batch. Each limb is divided into multiple fully-pipelined batches directed to the \textit{HP-IP}. The component enables parallel computation of 24 batches, involving 4 input limbs and 6 keys, resulting in significant parallelism for \texttt{IP} operation in \SolutionName.

\subsubsection{\textit{E-Key Buffer}}
The \textit{E-Key Buffer} is a storage unit that holds the evaluation key required for \texttt{IP} operations. It has a total size of \(2\beta\widetilde{\beta}\alpha^{'}\) polynomials in typical, which is equivalent to 189MB for \(N=2^{16}\). We have carefully designed the dataflow for the \textit{HP-IP} component to reduce memory size. By operating on 24 batches in parallel, the \textit{E-Key Buffer} now stores only the necessary essential batches for its computation. This adjustment, aligned with the \textit{HP-IP} component's requirements, allows \SolutionName~to allocate a 6.75MB \textit{E-Key Buffer} for each cluster
\footnote{
The \textit{E-Key Buffer} can store up to 6144~(256\(\times\)24 batches per computation) batches to accommodate pipeline overlap delays, enabling \textit{HP-IP} to perform 256 continuous computations.
}.


\subsection{Dataflow optimization for the KLSS \texttt{KeySwitch}}
\label{sec:design:C}


\begin{figure}[t]
    \centering
    \includegraphics[width=0.8\linewidth]{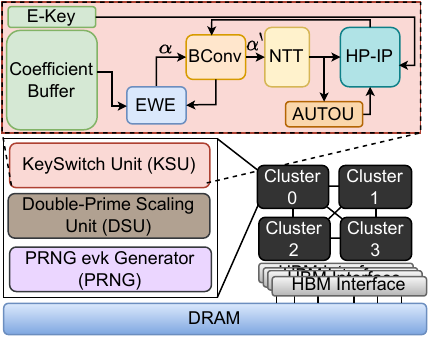}
    \vspace{-10pt}
    \caption{The overall architecture of \SolutionName~and the dataflow for the \texttt{KeySwitch} operation. The AUTOU, DSU, and PRNG components are consistent with those in SHARP\cite{sharp} \RebuttalChange{DSU executes the Double Scale operation to achieve more high precision~\cite{sharp,agrawal2023high}, PRNG generates the key required by the HPIP~\cite{craterlake,sharp}.}}
    \vspace{-15pt}
    \label{fig:overallksw}
\end{figure}

\RebuttalChange{Based on the dataflow illustrated in Figure~\ref{fig:kswnew}, \SolutionName~integrates specialized components for executing the KLSS-based \texttt{KeySwitch} operation, thereby directly enhancing the performance of FHE applications.}
To optimize these components, we propose an efficient data-flow scheme (Figure~\ref{fig:overallksw}). 
\RebuttalChange{For the KeySwitch operation within the \texttt{HMULT} operation, the accelerator follows the computation pattern of \texttt{BConvU} \(\rightarrow\) \texttt{NTTU} \(\rightarrow\) \texttt{HP-IP}~(To support Figure~\ref{fig:kswnew}(a),(b)). Additionally, during the execution of the BSGS operation in bootstrapping, the accelerator selects the datapath of \texttt{AUTOU} to execute the computation pattern of \texttt{BConvU} \(\rightarrow\) \texttt{NTTU} \(\rightarrow\) \texttt{AUTOU} \(\rightarrow\) \texttt{HP-IP}. This dataflow significantly improves the performance of BSGS~\cite{mad}. Additionally, other hardware units schedule corresponding components based on the software stack's execution and retrieve data from the relevant storage units.}

\subsubsection{Reduction of \texttt{BConv} operation}

As outlined in Sections~\ref{sec:motiv:a} and~\ref{sec:motiv:c}, the KLSS method exhibits an increased computational load in the \texttt{BConv} operation compared to the previous method. To address this added overhead, optimization techniques are deemed necessary.

A detailed analysis of the KLSS method's workflow reveals that the \textit{Recover-Limbs} stage significantly contributes to the augmented \texttt{BConv} operation. In this stage, a total of \(\widetilde{\beta}\) \texttt{BConv} operations are conducted to convert limbs from \(R^{T}\) to \(R^{PQ}\). For the last group, two consecutive \texttt{BConv} operations are performed on the limbs: the first to transition from \(R^{T}\) to \(R^{P}\) and the second for the transform to \(R^{Q}\) for the \texttt{ModDown} operation. Consequently, we propose a direct conversion of the last group's limbs from \(R^{T}\) to \(R^{Q}\) to optimize the execution of the \texttt{ModDown} operation.

\begin{equation}
\label{bconvreduce}
\begin{aligned}
     a_{R_{P}} &\approx \texttt{BConv}_{T\rightarrow P}(a_{R^{T}}) = a_{R^{P}}+Te_{1} \\
     a_{R_{Q}} &\approx \texttt{BConv}_{P\rightarrow Q}(a_{R^{P}}) = a_{R^{Q}}+Te_{1} + Pe_{2} \\ 
     \implies& a_{R_{Q}} \approx \texttt{BConv}_{T\rightarrow P}(a_{R^{T}}) = a_{R^{Q}}+Te \\
\end{aligned}
\end{equation}
\textbf{Noise reduction:} The \texttt{BConv} operation is an approximate scheme based on the Chinese Remainder Theorem~(CRT), which introduces noise bias once for each transforms~\cite{ckks}. As depicted in Eq.~\ref{bconvreduce}, the conversion from \(R^{T}\) to \(R^{P}\) leads to the introduction of T-fold noise. Consequently, two consecutive conversions result in the accumulation of two types of noise, \(R^{T}\) and \(R^{P}\).
Hence, as shown in Eq.~\ref{bconvreduce}, we propose to streamline the operation for the \textit{Recover-Limbs} stage, aiming to mitigate the noise caused by the \texttt{BConv} operation.

\label{sec:bconvfusion}

\subsubsection{\texttt{PtMatVecMult} optimization}


The \texttt{PtMatVecMult} operation, pivotal in \texttt{Bootstrapping}, undergoes optimization via the KLSS method, employed in \textit{CoeffToSolt} and \textit{SoltToCoeff} operations. The operation applies \texttt{BConv} to transform limbs into \(R^{T}\), followed by \(r\) sets of operation groups, each comprising two \texttt{Automorph} and one \texttt{IP} operation. Limbs then undergo \texttt{ModDown} to reduce the multiplicative level.

Through these optimizations and the designed workflow for \SolutionName, we achieve high performance in KLSS-based FHE applications. Strategies focus on streamlining batch-level data flow and minimizing redundant NTT operations, resulting in exceptional performance in executing applications on \SolutionName.

\begin{figure}[h]
    \centering
    \includegraphics[width=0.9\linewidth]{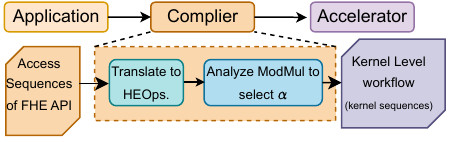}
    \vspace{-10pt}
    \caption{
    \RebuttalChange{
    The overall workflow of the Compiler-assisted optimization for the FHE applications.}}
    \vspace{-15pt}
    \label{fig:alpha}
\end{figure}

\subsection{Compiler-assisted optimization for \(\alpha\) for FHE applications}\label{sec:design:d}


We develop a compiler for CKKS FHE applications to optimize performance for varying multiplicative levels.
\RebuttalChange{
As shown in Figure~\ref{fig:alpha}, this compiler operates in two steps: 
1) It translates requests from applications into FHE operations. 2) It analyzes the number of ModMul times, selects \(\alpha\) values for each level based on N, and generates kernel sequences associated with distinct \(\alpha\) values for different levels, as explained in Section~\ref{sec:motiv:d}. Therefore, when a ciphertext undergoes a \texttt{keyswitch} operation, the alpha value is selected based on the limb count without additional switching time.
}


\begin{figure*}[t]
    \centering
    \includegraphics[width=\linewidth]{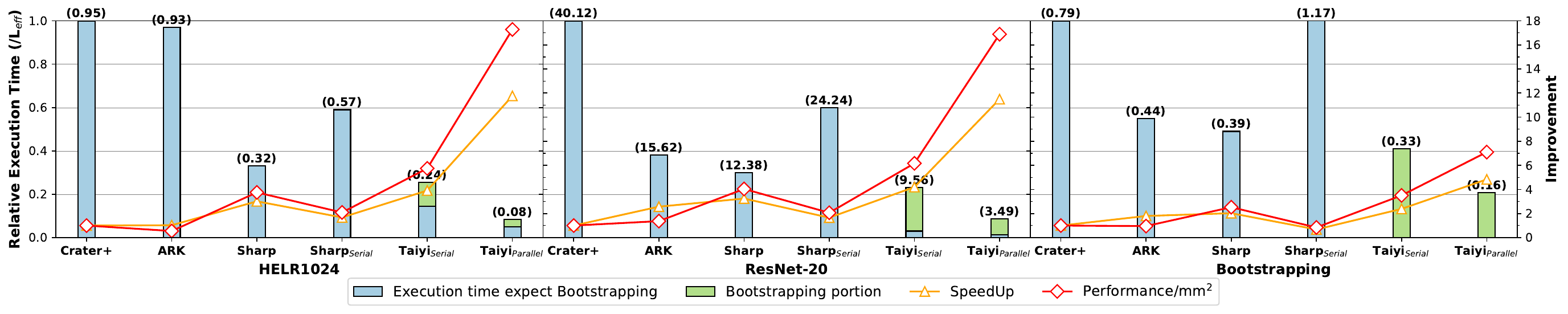}
    \vspace{-25pt}
    \caption{\RebuttalChange{Performance Comparison of \SolutionName~with Prior Accelerators. Evaluation within HELR1024, ResNet-20, and Bootstrapping (Normalized by \(\text{L}_{\text{eff}}\))}}
    \label{fig:finall}
    \vspace{-10pt}
\end{figure*}

\section{Methodology}

\label{sec:meth}

We implement all components of \SolutionName~in Verilog using the TSMC 7nm Processing Design Kit~(PDK)\cite{liu20211}. These components are designed with full pipelining and operate at a frequency of 1GHz. Additionally, we have incorporated the SRAM components provided by the PDK.

We constructed a cycle-level simulator based on the SimFHE~\cite{mad} for \SolutionName, consisting of two core components: the \textit{Instruction Generator} and the \textit{Hardware Executor}. The former executes FHE applications, generating a sequence of HE operations. These operations are then transformed into a data dependence graph of \textit{Arithmetic} kernels, subsequently sent to the \textit{Hardware Executor} for hardware behavior simulation.
\RebuttalChange{The \textit{Hardware Executor} selects appropriate components based on the calculation graph and tracks the frequency of their execution by processing the data flow received from the \textit{Instruction Generator}. Concurrently, it calculates the time needed to execute instructions based on the pipeline depth of the components and the total number of instructions.}

\begin{table}[t]
    \centering
        \caption{\RebuttalChange{CKKS instance parameter used for evaluation.}}
    \begin{tabular}{c|ccccccc}
    \hline
         Parameter & N& L& \(d_{\text{num}}\)& \(\alpha\) & \(\alpha^{'}\) &\(\widetilde{\beta}\) &\(L_{\text{eff}}\)   \\ \hline
         Value&\(2^{16}\) &38& 6& Dynamic& 4& \(\frac{L+\alpha}{\alpha}\)& 17 \\ \hline 
    \end{tabular}
    \vspace{-20pt}
    \label{tab:ckkspara}
\end{table}

We conducted runtime measurements for bootstrapping and representative FHE CKKS workloads by the simulator. To optimize performance and maintain a 128-bit security level, we applied KLSS methods~\cite{newkey}. 
\RebuttalChange{As demonstrated in Table~\ref{tab:ckkspara}, through the collaborative analysis of the new algorithm and hardware in Section~\ref{sec:motiv:b}, we have identified parameters that can efficiently execute FHE applications.}
The utilized open-sourced workloads are:
\begin{itemize}
\item \textbf{Bootstrapping~\cite{mouchet2020lattigo}:} This example involves performing bootstrapping operations, which consumes the multiplicative \(L_{\text{boot}} = 21\)~\cite{mouchet2020lattigo}.  
\item \textbf{HELR1024~\cite{helr,mad}:} For this application, we trained a logistic regression (LR) model on binary classification data using the FHE CKKS scheme~\cite{helr}. 
\item \textbf{ResNet-20~\cite{res20}:} We employed the FHE CKKS implementation of the ResNet-20~\cite{lee2022privacy} model from~\cite{res20}, applying it to 32 $\times$ 32 $\times$ 3 CIFAR-10~\cite{res20} images for once CNN inference.
\end{itemize}

\RebuttalChange{The selected applications encompass a variety of tasks, effectively demonstrating the versatility and functionality of our proposed ASIC accelerator. To provide a more detailed illustration of \SolutionName's performance, we simulate both serial and parallel modes as follows.
\begin{itemize}
    \item \(\text{Sharp}_\text{Serial}\): Implements the optimization methods described in this paper on Sharp~\cite{sharp}, with each component executing sequentially and without overlap.
    \item \(\text{\SolutionName}_\text{Serial}\): Implements the optimization methods outlined in this paper on \SolutionName, with each component executing sequentially and without any overlap.
    \item \(\text{\SolutionName}_\text{Parallel}\): Implement the optimization method outlined in this article on \SolutionName, where computing components can be executed concurrently.
\end{itemize}
In the serial mode, the accelerator's components execute sequentially, with only one component running in each cycle. In contrast, the parallel mode allows all components to execute simultaneously on the accelerator. For example, HPIP can process the output of NTTU concurrently with NTTU processing other input data.
}

\section{Results}










\subsection{Performance and Efficiency Analysis of \SolutionName}

\begin{table}[t]
    \centering
    \caption{Resource parameter in FHE Accelerators: Bandwidth (BW) and Capacity (Cap) Metrics. All accelerator working at 1GHz Operation Frequency .}\label{tab:main_res}
    \vspace{-5pt}
    \begin{tabular}{cccccc}\hline
    &Crater+&BTS&ARK&Sharp&\SolutionName\\ \hline
    \# of lanes& 2048&  2048& 1024 &1024 & 1024\\

    \multirow{1}{*}{On-chip} & \multirow{2}{*}{256}&  \multirow{2}{*}{512} &   \multirow{2}{*}{512}&  \multirow{2}{*}{180}&  \multirow{2}{*}{128.25}\\
     Mem Cap~(MB)&&&&& \\
     \multirow{1}{*}{On-chip} & \multirow{2}{*}{84}&  \multirow{2}{*}{-} &   \multirow{2}{*}{92}&  \multirow{2}{*}{72}&  \multirow{2}{*}{75}\\
     Mem BW~(TB/S)&&&&& \\
     Power~(W) &-&163.2&281&98&80    \\
     Area~(\(\text{mm}^{2}\))& 222.7&  373.6& 418.3 &178.8 & 151.6
    \\ \hline
    \end{tabular}
     \vspace{-10pt}
\end{table}

This section comprehensively explains the optimization approach to enhance FHE application performances. The performance evaluation of \SolutionName~is conducted, including execution time and chip area, compared to previous designs. Table~\ref{tab:main_res} illustrates that \SolutionName ~outperforms Crater+, BTS, ARK, and Sharp, exhibiting reductions of \Tofill{31.9\%}, \Tofill{59.4\%}, and \Tofill{63.7\%}, and \Tofill{15.7\%} in chip area, respectively. 
Furthermore, \SolutionName~attains bandwidth reductions of \Tofill{1.12\(\times\)} and \Tofill{1.22\(\times\)} in comparison to Crater+ and ARK, respectively. Compared to Sharp, \SolutionName~experiences only a 4.16\% increase in bandwidth requirements.
Besides, we also reduce \Tofill{28.75\%} on-chip memory requirement compared to the previous minimum accelerator.




\begin{figure*}[ht]
    \centering
    \begin{minipage}{0.52\textwidth}
        \centering
        \includegraphics[width=\linewidth]{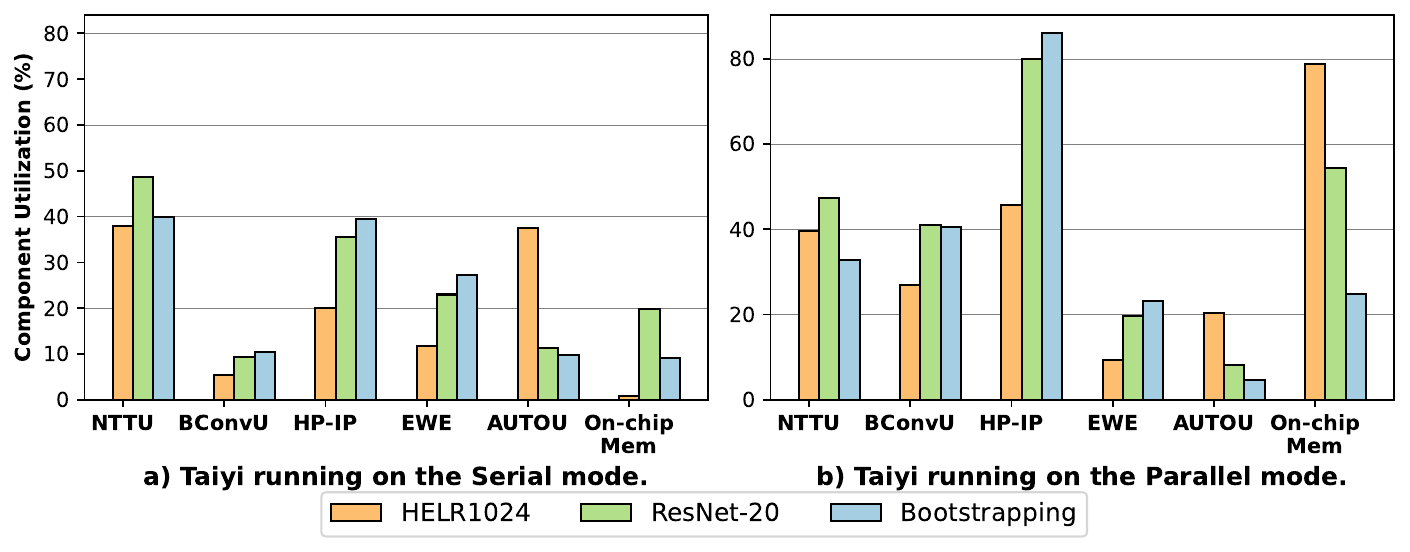}
        \vspace{-20pt}
        \caption{Utilization of \textit{SolutionName} Components.}
        \label{fig:utilandere}
    \end{minipage}
    \begin{minipage}{0.4\textwidth}
        \centering
        \includegraphics[width=\linewidth]{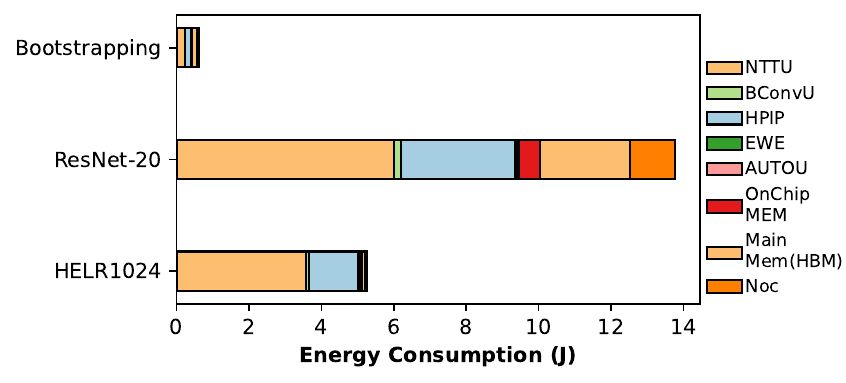}
        \caption{Energy Consumption of \textit{SolutionName} Components.}
        \label{fig:energy}
    \end{minipage}
    \vspace{-10pt}
\end{figure*}

\RebuttalChange{
As shown in Table~\ref{tab:absexetime}, \SolutionName~achieves an average performance improvement of \Tofill{1.17\(\times\)} in serial mode compared to \(\text{Sharp}_\text{Serial}\). This result indicates that the optimized hardware design efficiently executes FHE applications without algorithm optimization, demonstrating that the optimized \texttt{NTTU} and designed \texttt{HP-IP} components enhance FHE application execution efficiency. Additionally, \(\text{\SolutionName}_\text{Parallel}\)~achieves \Tofill{2.83\(\times\)} and \Tofill{1.54\(\times\)} performance improvement compared to \(\text{\SolutionName}_\text{Serial}\) and the previous state-of-the-art (SOTA)~\cite{sharp}. This achievement suggests that \SolutionName~exhibits strong concurrency, capable of executing multiple components simultaneously. Hardware and algorithm optimizations reduce the execution time of the \texttt{KeySwitch} operation by optimizing \(\alpha\) and decreasing the execution times of \texttt{NTT} and \texttt{BConv} operations.
}

\begin{table}[t]
    \centering
    \caption{\RebuttalChange{Comparison of Absolute performance between \(Taiyi\)  and other prior works~(ms).}}
    \begin{tabular}{ccccc}
    \hline
 &HELR1024&ResNet-20 & Bootstrapping& \(\text{L}_{\text{eff}}\) \\ \hline
 CPU & 23.3(s)& 2271(s)& 17.2(s)&8\\
Crater+&7.62&	321&	6.32&8\\
ARK&7.42&	125&	3.52&8 \\ 
Sharp&2.53&	99&	3.12&8\\
\(\text{Sharp}_{\text{Serial}}\)&4.54&	193.89&	9.35&8\\
\(\text{Taiyi}_{\text{Serial}}\)&4.14&	162.60&	7.61&17\\
\(\text{Taiyi}_{\text{Parallel}}\)&\textbf{1.37}&	\textbf{59.33}&	\textbf{2.77}&17\\ \hline
    \end{tabular}
    \label{tab:absexetime}
    \vspace{-20pt}
\end{table}

\RebuttalChange{
As indicated in Table~\ref{tab:absexetime}, our design exhibits higher \(L_\text{eff}\) after security analysis, demonstrating its ability to perform more effective multiplications for various FHE applications. Therefore, as illustrated in Figure~\ref{fig:finall}, we further evaluate the efficiency of application execution time. A lower allocation time signifies more effective execution of FHE operations, suggesting that our design can achieve better execution efficiency for a wider range of FHE applications. \SolutionName~achieves a performance improvement of \Tofill{7.04\(\times\)} compared to \(\text{Sharp}_{\text{Serial}}\). Besides, it achieves a \Tofill{68.1\%} performance improvement compared to the results reported in Sharp~\cite{sharp}.}

\RebuttalChange{
We conducted a comparison of performance per area with previous designs. The results show that our design can achieve an average improvement of \Tofill{13.75\(\times\)}, \Tofill{17.13\(\times\)}, \Tofill{3.87\(\times\)}, and \Tofill{8.31\(\times\)} for Crater+, ARK, Sharp, and \(\text{Sharp}_{\text{Serial}}\).}

Figure~\ref{fig:finall} illustrates that the \texttt{Bootstrapping} operation constitutes the primary overhead in FHE applications, accounting for \Tofill{84.3\%} and \Tofill{40.7\%} for ResNet-20 and HELR1024, respectively. Hence, it remains the principal source of overhead in the KLSS-based method.



\subsection{Component Utilization and Energy Consumption Analysis}
\label{sec:res:b}

This section focuses on the computation components (NTTU, BConvU, HP-IP, EWE, and AutoU) for analyzing utilization and energy consumption per area. As depicted in Figure~\ref{fig:utilandere}, for the serial mode, \texttt{HP-IP} emerges as the most utilized component in \texttt{Bootstrapping} applications, achieving 39.4\%. The utilization of NTTU, HP-IP, and EWE in all applications is substantial, averaging 31.47\%. This observation indicates that our design and optimization efforts achieve a balanced utilization across all components in executing FHE applications. \RebuttalChange{In the parallel mode, the \texttt{HP-IP} achieves 85.94\% utilization for the \texttt{Bootstrapping} operation and an average utilization of 70.45\% across the above FHE applications. This indicates that the \texttt{HP-IP} can be efficiently utilized during the execution of FHE applications. Additionally, the other components also achieve a balanced utilization during application execution.}

Furthermore, we provide energy consumption data for applications on \SolutionName, demonstrating that \SolutionName~consumes low power. As shown in Figure~\ref{fig:utilandere}, ResNet-20 only consumes 5.99J on NTT, with \texttt{HP-IP}
consuming 3.15J. These results affirm that \SolutionName~is a low-power accelerator, dissipating less than 80W of power for all FHE applications.

\begin{figure}[b]
    \centering
    \vspace{-10pt}/
    \includegraphics[width=0.5\linewidth]{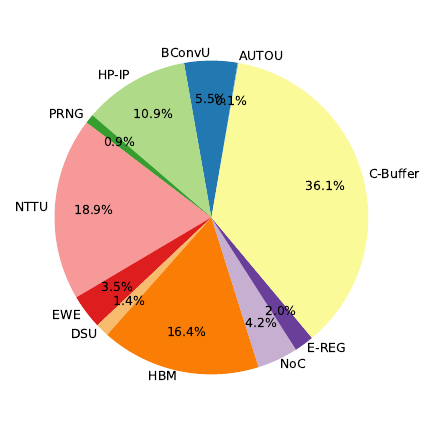}
    \vspace{-20pt}
    \caption{\RebuttalChange{Breakdown of accelerator area, indicating the distribution of resources among key components: \texttt{C-Buffer}(36.1\%), \texttt{HP-IP} (10.9\%), \texttt{NTTU} (18.9\%), \texttt{BConv} (5.5), and \texttt{Others} (28.6\%).}}
    \vspace{-10PT}
    \label{fig:area}
\end{figure}

\subsection{Accelerator area breakdown}
\label{sec:res:c}

\RebuttalChange{
As depicted in Figure~\ref{fig:area}, the predominant area within the accelerator is occupied by storage-related components, comprising up to 38.15\%. This observation underscores the substantial storage demands inherent in the FHE accelerator hardware design. If only increasing EWE of sharp~\cite{sharp} will significantly increase the chip area. Therefore, the \texttt{HP-IP} component represents a modest 10.9\% of the total area, resulting in a significant 74.2\% reduction in area overhead compared to equivalent EWE units with similar parallelism. This efficiency is achieved through the shared utilization of compute units for the same limbs within the \texttt{HP-IP}.}

\subsection{Sensitivity Study}
\subsubsection{Impact of \(N\) on Optimal \texttt{NTT} Operation}
\begin{figure}[t]
    \centering
    \includegraphics[width=\linewidth]{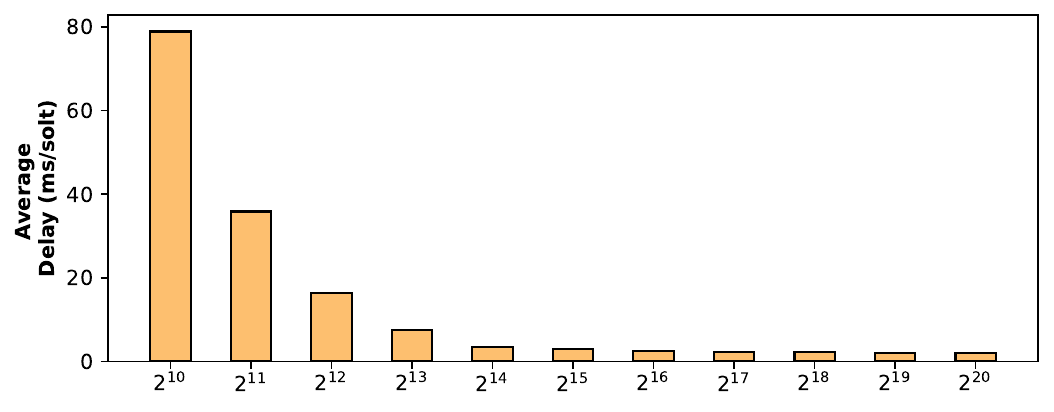}
    \vspace{-20pt}
    \caption{Execution Time per solt for different \(N\) within \SolutionName.}
    \vspace{-15pt}
    \label{fig:sent1}
\end{figure}

In this section, to comprehensively assess the performance of NTTU in \SolutionName, we conducted evaluations across polynomials of varying degrees. We analyzed the average computation delay for each modulus multiplication. As depicted in Figure~\ref{fig:sent1}, the average computation time decreases with increased polynomial length. However, when the length reaches \(2^{16}\), the rate of descent begins to slow down, suggesting that our design effectively supports calculations with polynomial lengths of \(2^{16}\) and maintains robust computational performance for longer polynomials.

\subsubsection{Impact of \(\alpha\) on Optimal \texttt{BConv} Operation}
\begin{figure}[t]
    \centering
    \includegraphics[width=0.9\linewidth]{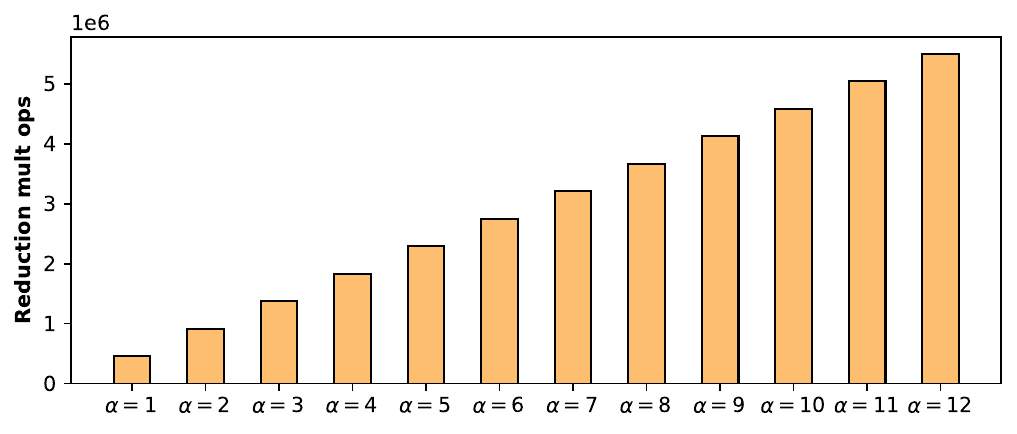}
    \vspace{-10pt}
    \caption{Multiplication operation number reduction with different \(\alpha\).}
    \label{fig:sent2}
    \vspace{-15pt}
\end{figure}

In this section, we conducted evaluations with varying \(\alpha\) values to analyze the computational performance of the optimized algorithm. As illustrated in Figure~\ref{fig:sent2}, it is evident that as \(\alpha\) increases, the reduction in computational workload also increases. Consequently, we can infer that the optimization algorithm proposed in this work is versatile, supporting any length of \(\alpha\) and effectively reducing computational complexity.

\section{Related Work}

 \subsection{GPU and FPGA acceleration method for FHE}
Numerous research efforts~\cite{100x, tensorfhe, fab, poseidon, wang2023he,shivdikar2023gme} have aimed to accelerate FHE applications on GPU and FPGA platforms. In the pioneering work by Ref.~\cite{100x}, FHE CKKS received initial GPU support, demonstrating a significant reduction in off-chip memory access through kernel fusion across primary functions. The work by Ref.~\cite{tensorfhe} introduced the use of tensor cores in recent NVIDIA GPUs for NTT acceleration, employing an approach that breaks down each 32-bit word into 8-bit integers fed into tensor cores. Despite this innovative technique, GPUs' limited on-chip memory capacity still results in frequent off-chip memory accesses, leading to memory-bound FHE workloads. Addressing this challenge, Ref.~\cite{fab} optimized the dataflow of the \texttt{KeySwitch} operation on FPGA, effectively overcoming the memory bandwidth bottleneck. Additionally, Ref.~\cite{poseidon} achieved notable performance enhancements for FHE applications through algorithm optimizations and modular fusion. Further contributions from research efforts like Ref.~\cite{tfhefpga, ye2022fpga, gener2021fpga} have accelerated application performance, particularly focusing on the TFHE scheme and deploying dedicated circuit designs for NTT/FFT operations. Despite the significant strides in these endeavors to accelerate FHE, a substantial performance gap persists, hindering the development of FHE.

 \subsection{ASIC accelerator design for FHE}
Several domain-specific accelerators have been proposed to enhance the practical performance of FHE applications, as detailed in works such as \cite{f1, bts, craterlake, ark, sharp, mad}. The first ASIC implementation supporting the CKKS scheme is presented in Ref.~\cite{f1}. However, it exhibits a performance gap in executing FHE applications and lacks support for the bootstrapping operation. Contrarily, \cite{bts, craterlake} optimize FHE applications through parameter selection, achieving comparable performance for bootstrapping and becoming the first to execute applications with unbounded multiplicative depth. \cite{ark} focuses on optimizing the BSGS operation using bootstrapping, significantly reducing the memory requirement of the bootstrapping operation. The state-of-the-art accelerator implementation for the CKKS scheme is Ref.~\cite{sharp}, which reduces on-chip memory size by decreasing the word length from the precision of real-world applications. It achieves optimal performance by enhancing the parallelism of the \texttt{BConv} component. Another optimization effort is presented by Ref.~\cite{mad}, which optimizes the dataflow to reduce on-chip memory requirements and also mitigates \texttt{NTT} operations. Additionally, Ref.~\cite{jiang2022matcha, putra2023strix,van2022fpt} contribute to accelerating TFHE applications by improving the performance of the blind-rotate operation through parallelization. Despite these notable acceleration efforts, these works still encounter performance bottlenecks for the KLSS method.

\section{Conclusion}

This paper introduces the \SolutionName, an accelerator specifically designed for KLSS-based FHE applications. We comprehensively analyze the computational workload differences between the KLSS-based KeySwitch method and the previous method. Our accelerator features dedicated hardware for the \texttt{Inner-product} operation, which ensures high performance while minimizing on-chip bandwidth requirements. Algorithmic optimizations are proposed to reduce the computational burden of the BConv operation. Additionally, we present a static compiler designed to dynamically select the optimal \(\alpha\) during the runtime of FHE applications.

\SolutionName~demonstrates high efficiency compared to state-of-the-art ASIC FHE accelerators designed for KLSS-based methods, showcasing a \RebuttalChange{3.8\(\times\)} enhancement in performance per area. This emphasizes the effectiveness of our design in addressing the specific computational challenges associated with KLSS-based FHE applications.



\bibliographystyle{IEEEtranS}
\bibliography{refs}

\end{document}